\documentclass[letterpaper, 10pt, conference]{ieeeconf}  

\IEEEoverridecommandlockouts                              

\usepackage{cite}
\usepackage{amsmath,amssymb,amsfonts}
\usepackage{algorithmic}
\usepackage{graphicx}
\usepackage{textcomp}

\usepackage{titlesec}
\usepackage{mathabx}
\usepackage{xcolor}
\usepackage{float}
\usepackage{multirow}
\usepackage{tikz}
\usetikzlibrary{shadows}

\usepackage{graphicx}
\usepackage{subcaption}
\usepackage{epsfig} 
\usepackage{cite}
\usepackage{ ACSD-CR }

\usepackage{mathrsfs}  
\DeclareMathAlphabet{\mathpzc}{T1}{pzc}{m}{it} 
\usepackage{bm} 
\usepackage{amsmath} 
\usepackage{amssymb} 
\usepackage{arydshln}
\usepackage{mathdots} 
\usepackage{xcolor}
\usepackage{braket}
\usepackage{algorithmic}
\usepackage{algorithm}
\restylefloat{algorithm} 

\newcommand{\R}{{\mathbb{R}}}         
\newcommand{\Z}{{\mathbb{Z}}}         
\newcommand{\N}{{\mathbb{N}}}         

\renewcommand{\S}{\mathbb{S}}          

\newcommand{\I}{\mathbb{I}}

\renewcommand{\c}[1]{\mathcal{#1}}
\newcommand{\Ac}{\c{A}}
\newcommand{\Bc}{\c{B}}
\newcommand{\Cc}{\c{C}}
\newcommand{\Dc}{\c{D}}

\newcommand{\Kc}{\c{K}}
\newcommand{\Lc}{\c{L}}

\newcommand{\Del}{\Delta}

\newcommand{\eps}{\varepsilon}

\newcommand{\ov}[1]{\overline{#1}}

\newcommand{\ti}[1]{\widetilde{#1}}

\newcommand{\wh}[1]{\widehat{#1}}

\newcommand{\U}{\mathbb{U}}
\newcommand{\X}{\mathbb{X}}
\newcommand{\Y}{\mathbb{Y}}

\newcommand{\W}{\mathbb{W}}

\makeatletter
\newcommand\RedeclareMathOperator{%
	\@ifstar{\def\rmo@s{m}\rmo@redeclare}{\def\rmo@s{o}\rmo@redeclare}%
}
\newcommand\rmo@redeclare[2]{%
	\begingroup \escapechar\m@ne\xdef\@gtempa{{\string#1}}\endgroup
	\expandafter\@ifundefined\@gtempa
	{\@latex@error{\noexpand#1undefined}\@ehc}%
	\relax
	\expandafter\rmo@declmathop\rmo@s{#1}{#2}}
\newcommand\rmo@declmathop[3]{%
	\DeclareRobustCommand{#2}{\qopname\newmcodes@#1{#3}}%
}
\@onlypreamble\RedeclareMathOperator

\RedeclareMathOperator{\Im}{Im} 
\RedeclareMathOperator{\Re}{Re} 
\RedeclareMathOperator{\div}{div}  
\RedeclareMathOperator{\ker}{ker}   

\DeclareMathOperator*{\diag}{diag}      
\DeclareMathOperator*{\col}{col}    

\newcommand{\cge}{\succcurlyeq}
\newcommand{\cle}{\preccurlyeq}
\newcommand{\cg}{\succ} 

\newcommand{\subs}{\subset}

\newcommand{\subeq}{\subseteq}

\newcommand{\mat}[2]{\left(\begin{array}{#1}#2\end{array}\right)}

\newcommand{\smat}[1]{\left(\begin{smallmatrix}#1\end{smallmatrix}\right)}

\newcommand{\as}[1]{\renewcommand{\arraystretch}{#1}} 
\newcommand{\mats}[2]{\scriptsize
	\as{1}
	\arraycolsep=1pt \mat{#1}{#2}}

\newcommand{\aln}[1]{\noindent\begin{align*}#1\end{align*}}
\newcommand{\all}[1]{\noindent\begin{align}#1\end{align}}
\newcommand{\mun}[1]{\vspace*{-1ex}\noindent
	\begin{multline*}#1\end{multline*}}

\newcommand{\equ}[1]{\begin{equation*}#1\end{equation*}}
\newcommand{\eql}[1]{\begin{equation}#1\end{equation}}
\newcommand{\subeql}[2]{\begin{subequations}{#1}\all{#2}\end{subequations}}

\newcommand{\eqa}[2]{\eql{#1\begin{aligned}#2\end{aligned}}}

\newtheorem{theo}{Theorem}[section]
\newtheorem{lemm}[theo]{Lemma}
\newtheorem{prpo}[theo]{Proposition}
\newtheorem{coro}[theo]{Corollary}
\newtheorem{defi}[theo]{Definition}
\newtheorem{rema}[theo]{Remark}
\newtheorem{assu}[theo]{Assumption}

\newenvironment{proofnew}[1][Proof of]{
	\it #1:  \rm}
{\hfill \footnotesize{$\blacksquare$}\vspace{2ex}}

\newcommand{\theorem}[1]{\begin{theo}#1\end{theo}}
\newcommand{\corollary}[1]{\begin{coro}#1\end{coro}}
\newcommand{\lemma}[1]{\begin{lemm}#1\end{lemm}}
\newcommand{\definition}[1]{\begin{defi}#1\end{defi}}

\newcommand{\remark}[1]{\begin{rema}#1\end{rema}}
\newcommand{\proo}[1]{\begin{proof}#1\end{proof}}
\newcommand{\prop}[1]{\begin{prpo}#1\end{prpo}}

\newcommand{\red}[1]{\textcolor{black}{#1}}
\newcommand{\Blu}[1]{\textcolor{blue}{#1}}
\newcommand{\todo}[1]{\textcolor{purple}{#1}}

\newcommand{\mam}[2]{\left\|\begin{array}{#1} #2 \end{array} \right\|}

\renewcommand{\todo}{}

\renewcommand{\Blu}[1]{\textcolor{black}{#1}}
\renewcommand{\todo}[1]{\textcolor{black}{#1}}

\begin{document}

\SETCR{\CRIEEE{DOI:10.1109/LCSYS.2025.3629926}{IEEE Control System Letters}{\title, \author }}

\title{MHE in Output Feedback Control of Uncertain Nonlinear Systems via IQCs (Extended Version)}

\author{
Yang Guo, 
Stefan Streif
\thanks{This work is supported by the European Social Fund Plus (ESF Plus) and the Free State of Saxony through the project HZwo:StabiGrid. }%
\thanks{The authors are with the Chemnitz University of Technology, Professorship  Automatic Control and System Dynamics, 
	09107 Chemnitz, Germany 
	\texttt{\{yang.guo, stefan.streif\}@etit.tu-chemnitz.de.}}
}

\maketitle
\thispagestyle{empty}

\begin{abstract}
We propose a moving horizon estimation (MHE) scheme for general nonlinear constrained systems with parametric or static nonlinear uncertainties and a predetermined state feedback controller that is assumed to robustly stabilize the system in the absence of estimation errors. Leveraging integral quadratic constraints (IQCs), we introduce a new notion of detectability that is robust to possibly non-parametric uncertainties and verifiable in practice. Assuming that the uncertain system driven by the controller satisfies this notion of detectability, we provide an MHE formulation such that the closed-loop system formed of the uncertain system, the controller and MHE is input-to-state stable w.r.t. exogenous disturbances.
\end{abstract}


\section{Introduction}

In many control applications and whenever the states can not be completely measured, state estimation is of paramount importance.   For nonlinear systems with bounded disturbances, states can be estimated via various approaches, such as Kazantzis-Kravaris/Luenberger observers \cite{Tran2024} and moving horizon estimation (MHE) \cite{ji2015,  knufer2023nonlinear, Schiller2023lyapunov}, just name a few.  \Blu{ The design framework  of MHE presented in \cite{Schiller2023lyapunov} is advanced in   \cite{schiller2023nonlinear}  and \cite{muntwiler2023mhe}  for robust nonlinear state estimation under  parametric uncertainties.} Estimator design  becomes particularly challenging in presence of non-parametric uncertainties, e.g., unmodeled nonlinearities and dynamics, which, in general, can not be  treated as bounded disturbances.   For linear time-invariant  systems  with norm-bounded non-parametric uncertainties, robust $H_\infty$ and $H_2$ estimators are developed in \cite{sun2005robust}. The work \cite{scherer2008} considers a larger   class of   uncertainties for linear systems   with less conservatism  by employing the framework of integral quadratic constraints (IQCs) \cite{megretski2002system, veenaman2016,scherer2022dissipativity}, which  allows dealing with  various classes of uncertainties.  For a class of  nonlinear systems with unmodeled dynamics,  \cite{zhang2015observer} proposes adaptive observers  by using  
dissipativity  under restrictive structural assumptions.

In the context of robust output feedback control using state estimates,     the \red{interaction between}  systems, controller and estimators \red{needs to be treated  carefully}, \red{especially} in the presence of  non-parametric uncertainties. 
 For  linear constrained systems with a norm-bounded uncertainty, a tube-based model predictive controller (MPC) combined with a linear  observer is proposed in \cite{lovaas2008robust} to ensure robust closed-loop stability.     The work \cite{schwenkel2025output} utilizes IQCs  to design a linear observer  for the output feedback MPC,   which is  robust to  larger classes of non-parametric uncertainties with less conservatism than the one in   \cite{lovaas2008robust}. 
 For strict-feedback nonlinear systems with dynamic uncertainties and unmodeled nonlinearities, \cite{tong2009combined} develops an adaptive  fuzzy output-feedback controller using a fuzzy state observer to ensure input-to-state practical stability.    
However, to the best of the authors’ knowledge, state estimation in output feedback control of general  nonlinear  systems with non-parametric uncertainties is still open. \red{Furthermore, in the above works, the estimator is designed prior to the controller and closed-loop stability is explicitly considered only in the design of controllers}   

\emph{Contributions and outline:} 
\red{We present an MHE framework for an output feedback control setup (cf. Figure~\ref{fig:closed_loop} in Section~\ref{sec:setup}) comprising a general nonlinear constrained system with a possibly nonlinear uncertainty and a predetermined feedback controller using state estimates. The controller  is supposed to be input-to-state stabilizing without estimation errors. To deal with uncertainties, we propose a notion of robust detectability by exploiting tailored IQCs \Blu{in Section~\ref{sec:detect}} and provide linear matrix inequality (LMI) conditions for the verification of this notion in Section~\ref{sec:verifiation}. Assuming that the system with the controller is robustly detectable, we present the main result in Section~\ref{sec:MHE} and formulate the MHE and show that the closed-loop system remains input-to-state stable (ISS) w.r.t. exogenous disturbances despite uncertainties. We exemplify the theoretical findings by a numerical example in Section~\ref{sec:example} and summarize the presented results in Section~\ref{sec:conclusion}.}

\emph{Notation:} \red{Component-wise vector inequalities are denoted by  $\leq, \geq$.} The set of non-negative integers (in $[a,b]$) is denoted by  $\N_0$ ($\I_{[a,b]}$).
The set of symmetric matrices in $\R^{n\times n}$ is denoted by $\S^ {n}$. The notions $0_{n\times m}$ and $I_{n}$  denote a  zeros  matrix in $\R^{n \times m}$ and an identity matrix in $\S^n$ respectively.   Given $P \cge 0$,  $\| x \|^2_P$  denotes  $x^\top P x$. For $A \cge 0$, $B \cg 0$,  $\ov{\lambda}(A,B)$ denotes the largest value $\lambda$ such that $\det(A-\lambda B)=0$. Further,   $T^\top A T$ in matrix inequalities \red{is abbreviated} by $(\bullet)^\top A T$. The symbols $\col(X_1, \ldots, X_N)$ and $\diag(X_1, \ldots, X_N)$ are used to  stack $X_1, \ldots, X_N$ vertically and diagonally respectively. 
  The class of continuous strictly increasing functions $\alpha: [0, \infty) \to [0, \infty)$ with $\alpha(0)=0$ is denoted by $\Kc$. The class of functions $\beta:[0, \infty) \times \N_0 \to [0, \infty)$ with $\beta(\cdot, k) \in \Kc$ for any fixed  $k \in \N_0$ and  non-increasing $\beta(r, \cdot)$ satisfying $\lim_{k \to \infty} \beta(r,k)=0$ for any fixed $r\in [0,\infty)$ is denoted by $\Kc \Lc$. 

\section{Problem Setup}
\label{sec:setup}
Let us consider the uncertain feedback interconnection  

\subeql{\label{eq:sys}}{
	x_{k+1} &= f(x_k, w_k, d_k, u_k), \label{eq:sys_1} \\
	y_k &= h(x_k, w_k, d_k, u_k),  \label{eq:sys_2}\\
	v_k &= \Blu{g(x_k, w_k)}, \label{eq:sys_3} \\
	d_k &= \Del(v_k),  \label{eq:sys_4}
}
involving  a  \Blu{known nonlinear system  \eqref{eq:sys_1}-\eqref{eq:sys_3}}   and  a memory-less (possibly nonlinear) uncertainty  \Blu{$\Del: \R^q \to \R^{p}$} with $\Delta(0)=0$. Further,   $v_k \in \R^q$ and  $d_k \in \R^p$ are the unmeasurable auxiliary output and input respectively. In addition, $x_k \in \X \subeq \R^n$ is the unmeasured state,  $w_k\in \W \subs \R^{n_w}$ is  the bounded disturbance with $0 \in \W$,   $u_k \in \U \subeq \R^l $ and  $y_k \in  \Y \subeq \R^m$ are  the control input and the output measurement respectively.
Moreover,   the function $f$, $h$ and \Blu{$g$ are assumed to be Lipschitz continuous on $\X \times \W \times \R^p \times \U$ and $\X \times \W $ respectively. Further, we assume that $g(0,0)=0$.} Throughout the paper, we denote the domain of trajectories $\X \times \W \times \R^p \times \U \times \R^p \times \Y$ by $\Z$ \Blu{and assume it is a Cartesian product of  intervals}. 


The control input is determined by a predefined controller
\eql{\label{eq:control_input}
	u_k = \kappa(\wh{x}_k),  
} 
with   the state estimate $\wh{x}_{k} \in \X$ of the system~\eqref{eq:sys} and the Lipschitz continuous function $\kappa: \X \to \U$. The controller $\kappa$ is designed to ensure that the system~\eqref{eq:sys} with $u_k=\kappa(x_k)$ using the true  state $x_k$ is input-to-state stable (ISS),
  that is, there exist $ \wh{\beta} \in \Kc \Lc$ and $ \wh{\alpha} \in \Kc$ such that, for  any trajectory $(x_i,w_i,d_i,v_i,y_i,\kappa(x_i))^\infty_{i=0} \in \Z^\infty$ satisfying \eqref{eq:sys}, 
\eql{\label{eq:ISS_nominal}
	\|x_k\| \leq  \wh{\beta}( \|x_0\|, k ) + \wh{\alpha}(\max_{i\in \I_{[0,k-1]}} \|w_i\|) 
} 
holds for all $k\in \N_0$. 
 Such a controller could be, e.g., 
  a nonlinear optimal feedback controller\cite{glad1987robustness}.

\begin{figure}[h]


\tikzset {_p2fu5rrt9/.code = {\pgfsetadditionalshadetransform{ \pgftransformshift{\pgfpoint{0 bp } { 0 bp }  }  \pgftransformrotate{0 }  \pgftransformscale{2 }  }}}
\pgfdeclarehorizontalshading{_51uvezjb4}{150bp}{rgb(0bp)=(1,1,1);
	rgb(37.5bp)=(1,1,1);
	rgb(62.5bp)=(0.9,0.9,0.9);
	rgb(100bp)=(0.9,0.9,0.9)}


\tikzset {_zmiiqxjtt/.code = {\pgfsetadditionalshadetransform{ \pgftransformshift{\pgfpoint{0 bp } { 0 bp }  }  \pgftransformrotate{0 }  \pgftransformscale{2 }  }}}
\pgfdeclarehorizontalshading{_ihgkty1h6}{150bp}{rgb(0bp)=(1,1,1);
	rgb(37.5bp)=(1,1,1);
	rgb(62.5bp)=(0.9,0.9,0.9);
	rgb(100bp)=(0.9,0.9,0.9)}


\tikzset {_d4ndbmwkm/.code = {\pgfsetadditionalshadetransform{ \pgftransformshift{\pgfpoint{0 bp } { 0 bp }  }  \pgftransformrotate{0 }  \pgftransformscale{2 }  }}}
\pgfdeclarehorizontalshading{_atztfdbz4}{150bp}{rgb(0bp)=(1,1,1);
	rgb(37.5bp)=(1,1,1);
	rgb(62.5bp)=(0.9,0.9,0.9);
	rgb(100bp)=(0.9,0.9,0.9)}


\tikzset {_ag057w52n/.code = {\pgfsetadditionalshadetransform{ \pgftransformshift{\pgfpoint{0 bp } { 0 bp }  }  \pgftransformrotate{0 }  \pgftransformscale{2 }  }}}
\pgfdeclarehorizontalshading{_eb3gz67bg}{150bp}{rgb(0bp)=(1,1,1);
	rgb(37.5bp)=(1,1,1);
	rgb(62.5bp)=(0.9,0.9,0.9);
	rgb(100bp)=(0.9,0.9,0.9)}
\tikzset{every picture/.style={line width=0.75pt}} 

\centering

\begin{tikzpicture}[x=0.75pt,y=0.75pt,yscale=-1,xscale=1]
	
	\path  [shading=_51uvezjb4,_p2fu5rrt9][general shadow={fill={rgb, 255:red, 208; green, 208; blue, 208 }  ,shadow xshift=3pt,shadow yshift=-2.25pt, opacity=1 }] (92.29,137.69) .. controls (92.29,131) and (97.71,125.57) .. (104.4,125.57) -- (202.74,125.57) .. controls (209.43,125.57) and (214.86,131) .. (214.86,137.69) -- (214.86,174.03) .. controls (214.86,180.72) and (209.43,186.14) .. (202.74,186.14) -- (104.4,186.14) .. controls (97.71,186.14) and (92.29,180.72) .. (92.29,174.03) -- cycle ; 
	\draw   (92.29,137.69) .. controls (92.29,131) and (97.71,125.57) .. (104.4,125.57) -- (202.74,125.57) .. controls (209.43,125.57) and (214.86,131) .. (214.86,137.69) -- (214.86,174.03) .. controls (214.86,180.72) and (209.43,186.14) .. (202.74,186.14) -- (104.4,186.14) .. controls (97.71,186.14) and (92.29,180.72) .. (92.29,174.03) -- cycle ; 
	
	\draw [line width=0.75]    (90,139.5) -- (72,139.53) -- (72,104) -- (117.9,104) ;
	\draw [shift={(92,139.5)}, rotate = 179.93] [fill={rgb, 255:red, 0; green, 0; blue, 0 }  ][line width=0.08]  [draw opacity=0] (12,-3) -- (0,0) -- (12,3) -- cycle    ;
	\path  [shading=_ihgkty1h6,_zmiiqxjtt][general shadow={fill={rgb, 255:red, 208; green, 208; blue, 208 }  ,shadow xshift=3pt,shadow yshift=-2.25pt, opacity=1 }] (118,96.15) .. controls (118,93.33) and (120.28,91.05) .. (123.1,91.05) -- (182.9,91.05) .. controls (185.72,91.05) and (188,93.33) .. (188,96.15) -- (188,111.45) .. controls (188,114.27) and (185.72,116.55) .. (182.9,116.55) -- (123.1,116.55) .. controls (120.28,116.55) and (118,114.27) .. (118,111.45) -- cycle ; 
	\draw   (118,96.15) .. controls (118,93.33) and (120.28,91.05) .. (123.1,91.05) -- (182.9,91.05) .. controls (185.72,91.05) and (188,93.33) .. (188,96.15) -- (188,111.45) .. controls (188,114.27) and (185.72,116.55) .. (182.9,116.55) -- (123.1,116.55) .. controls (120.28,116.55) and (118,114.27) .. (118,111.45) -- cycle ; 
	
	\draw    (69.68,154.53) -- (90,154.5) ;
	\draw [shift={(92,154.5)}, rotate = 179.93] [fill={rgb, 255:red, 0; green, 0; blue, 0 }  ][line width=0.08]  [draw opacity=0] (12,-3) -- (0,0) -- (12,3) -- cycle    ;
	\draw [line width=0.75]    (215,139.5) -- (233,139.5) -- (233,104) -- (190,104) ;
	\draw [shift={(188,104)}, rotate = 360] [fill={rgb, 255:red, 0; green, 0; blue, 0 }  ][line width=0.08]  [draw opacity=0] (12,-3) -- (0,0) -- (12,3) -- cycle    ;
	\draw    (215,161.05) -- (247.5,161.03) ;
	\draw [shift={(249.5,161.03)}, rotate = 179.96] [fill={rgb, 255:red, 0; green, 0; blue, 0 }  ][line width=0.08]  [draw opacity=0] (12,-3) -- (0,0) -- (12,3) -- cycle    ;
	\path  [shading=_atztfdbz4,_d4ndbmwkm][general shadow={fill={rgb, 255:red, 208; green, 208; blue, 208 }  ,shadow xshift=3pt,shadow yshift=-2.25pt, opacity=1 }] (120.5,205.25) .. controls (120.5,202.59) and (122.66,200.43) .. (125.32,200.43) -- (170.85,200.43) .. controls (173.52,200.43) and (175.68,202.59) .. (175.68,205.25) -- (175.68,219.73) .. controls (175.68,222.39) and (173.52,224.55) .. (170.85,224.55) -- (125.32,224.55) .. controls (122.66,224.55) and (120.5,222.39) .. (120.5,219.73) -- cycle ; 
	\draw   (120.5,205.25) .. controls (120.5,202.59) and (122.66,200.43) .. (125.32,200.43) -- (170.85,200.43) .. controls (173.52,200.43) and (175.68,202.59) .. (175.68,205.25) -- (175.68,219.73) .. controls (175.68,222.39) and (173.52,224.55) .. (170.85,224.55) -- (125.32,224.55) .. controls (122.66,224.55) and (120.5,222.39) .. (120.5,219.73) -- cycle ; 
	
	\draw [line width=0.75]    (120.6,212) -- (72,212) -- (72,168) -- (90,168) ;
	\draw [shift={(92,168)}, rotate = 180] [fill={rgb, 255:red, 0; green, 0; blue, 0 }  ][line width=0.08]  [draw opacity=0] (12,-3) -- (0,0) -- (12,3) -- cycle    ;
	\path  [shading=_eb3gz67bg,_ag057w52n][general shadow={fill={rgb, 255:red, 208; green, 208; blue, 208 }  ,shadow xshift=3pt,shadow yshift=-2.25pt, opacity=1 }] (249.5,156.05) .. controls (249.5,151.63) and (253.08,148.05) .. (257.5,148.05) -- (296.68,148.05) .. controls (301.1,148.05) and (304.68,151.63) .. (304.68,156.05) -- (304.68,180.03) .. controls (304.68,184.45) and (301.1,188.03) .. (296.68,188.03) -- (257.5,188.03) .. controls (253.08,188.03) and (249.5,184.45) .. (249.5,180.03) -- cycle ; 
	\draw   (249.5,156.05) .. controls (249.5,151.63) and (253.08,148.05) .. (257.5,148.05) -- (296.68,148.05) .. controls (301.1,148.05) and (304.68,151.63) .. (304.68,156.05) -- (304.68,180.03) .. controls (304.68,184.45) and (301.1,188.03) .. (296.68,188.03) -- (257.5,188.03) .. controls (253.08,188.03) and (249.5,184.45) .. (249.5,180.03) -- cycle ; 
	
	\draw [line width=0.75]    (177.62,212) -- (278,212) -- (278,188) ;
	\draw [shift={(175.62,212)}, rotate = 0] [fill={rgb, 255:red, 0; green, 0; blue, 0 }  ][line width=0.08]  [draw opacity=0] (12,-3) -- (0,0) -- (12,3) -- cycle    ;
	\draw    (72,195.29) -- (225,195.29) -- (225,174.29) -- (247.5,174.29) ;
	\draw [shift={(249.5,174.29)}, rotate = 180] [fill={rgb, 255:red, 0; green, 0; blue, 0 }  ][line width=0.08]  [draw opacity=0] (12,-3) -- (0,0) -- (12,3) -- cycle    ;
	
	\draw (153.57,152) node   [align=left] {\begin{minipage}[lt]{73pt}\setlength\topsep{0pt}
			\begin{center}
			 {\small \aln{x^+&=f(x, w,d,u)\\y&=h(x, w,d,u)\\v&=\Blu{g(x,w)}  } }
			\end{center}
	\end{minipage}};
	\draw (146.33,97) node [anchor=north west][inner sep=0.75pt]    {$\Delta$};
	\draw (61.7,111.27) node    {$d$};
	\draw (61.7,152.2) node    {$w$};
	\draw (61.7,200) node    {$u$};
	\draw (242,111.3) node    {$v$};
	\draw (242,144.13) node    {$y$};
	\draw (260,162) node [anchor=north west][inner sep=0.75pt]   [align=left] {MHE};
	\draw (142,209) node [anchor=north west][inner sep=0.75pt]    {$\kappa$};
	\draw (290.75,200) node    {$\hat{x}$};
\end{tikzpicture}
	\vspace{2pt}
	\caption{ Interconnection of the uncertain system~\eqref{eq:sys}, i.e., \eqref{eq:sys_1}--\eqref{eq:sys_3} with uncertainty $\Delta$, MHE and \red{controller} $\kappa$. }
	\label{fig:closed_loop}	
\end{figure}
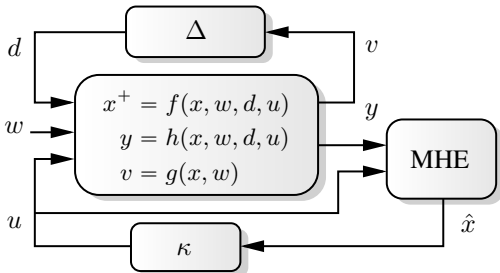

The goal is to design an MHE to estimate the state of   uncertain system~\eqref{eq:sys} controlled by \eqref{eq:control_input} such that the closed-loop system depicted in Fig.~\ref{fig:closed_loop} remains ISS, i.e. there exist $\wh{\beta} \in \Kc \Lc$ and $\wh{\alpha} \in \Kc$ such that
\eql{\label{eq:RGES}
	\|x_k\| \leq  \wh{\beta}( \|x_0\|+\|x_0-\wh{x}_0\|, k ) + \wh{\alpha}(\max_{i\in \I_{[0,k-1]}} \|w_i\|)
} 
holds for all $k\in \N_0$, any  $\wh{x}_0 \in \X$, and any trajectory $(x_i,w_i,d_i,v_i,y_i,\kappa(\wh{x}_i))^\infty_{i=0} \in \Z^\infty$ of the system \eqref{eq:sys}. 
 
\section{Robust Detectability with IQCs} 
\label{sec:detect}

In this section,   we introduce a concept of robust detectability, 
which  serves as a starting point  for  the design of a robust estimator  with   closed-loop stability. 
To this end, we   replace  $\Delta$ with  constraints on its inputs and outputs by means of  the following notion of IQCs. 
\definition{[Point-wise $\rho$--IQC]  \label{def:pointwiseIQC}
	The uncertainty  $\Del: \Blu{\R^{q}} \to \R^{p}$ satisfies point-wise $\rho$--IQC defined by $M\in \S^{n_z}, Z \in \S^{n_{\psi}}$, $(A_\Psi, B_\Psi, C_\Psi,D_\Psi)$ and $ \rho \in (0,1)$,  if
	\eql{ \label{eq:pointIQC}
		z_k^\top M z_k- \psi^\top_{k}Z \psi_{k} +  \rho^{-2} \psi^\top_{k+1} Z \psi_{k+1}  \geq 0 
	}
	for all $k\in \N_0$ and  every trajectory $(\psi_k, v_k, \Delta(v_k), z_k)^{\infty}_{k=0}$ of the filter $\Psi$ 
	\eqa{\label{eq:pointIQC_filter}}{
		\psi_{k+1} &= A_\Psi \psi_k +B_\Psi  \col(v_k, \Del(v_k)), \ \psi_0=0,  \\
		z_k &= C_\Psi \psi_k +D_\Psi  \col(v_k, \Del(v_k)).   
	}
}
\Blu{\remark{By multiplying \eqref{eq:pointIQC} by $\rho^{-2k}$ and 
 defining $\bar{z}_k:=\rho^{-k}z_k$,  $\bar{\psi}_k:=\rho^{-k}\psi_k$ and $\bar{v}_k:=\rho^{-k}v_k$, we can reformulate \eqref{eq:pointIQC} and \eqref{eq:pointIQC_filter} into
\eql{\label{eq:pointIQC_gen}
\bar{z}_k^\top M \bar{z}_k- \bar{\psi}^\top_{k}Z \bar{\psi}_{k} +   \bar{\psi}^\top_{k+1} Z \bar{\psi}_{k+1}  \geq 0 
}
and 
\aln{
\bar{\psi}_{k+1} &= \rho A_\Psi \bar{\psi}_k + \rho B_\Psi  \col(\bar{v}_k, \rho^{-k}\Del(\rho^{k} \bar{v}_k)), \ \bar{\psi}_0=0,  \\
		\bar{z}_k &= C_\Psi \bar{\psi}_k +D_\Psi  \col(\bar{v}_k, \rho^{-k}\Del(\rho^{k} \bar{v}_k)). 
}
This means that the weighted uncertainty $\rho^{-k}\circ\Delta\circ \rho^k$ satisfies the so-called point-wise IQC with storage defined in \cite[Theorem~2]{morato2023stabilizing}.
Summing \eqref{eq:pointIQC_gen} from $k=0$ to $L\in \N$ yields
\eql{
\sum^{L}_{k=0}\bar{z}_k^\top M \bar{z}_k +    \bar{\psi}^\top_{L+1} Z \bar{\psi}_{L+1}  \geq 0, 
}
which is the discrete-time version of finite-horizon IQC with terminal costs proposed in \cite{scherer2022dissipativity}. This formulation is strongly tied to classical frequency-domain IQC theory\cite{megretski2002system}.
It is straightforward to see that a weighted uncertainty satisfying  point-wise IQC with storage satisfies finite-horizon IQC with terminal costs trivially, but the reverse is not necessarily true. Therefore, the class of uncertainties, which can be described by IQC according to Definition~\ref{def:pointwiseIQC},  might be limited.    
}}


\definition{ [\todo{Robust Detectability}] \label{def:robust_detec} 
	The system~\eqref{eq:sys} with the controller~\eqref{eq:control_input} is robustly detectable if there exist $(M, Z, A_\Psi, B_\Psi, C_\Psi,D_\Psi, \rho)$  defining  point-wise $\rho$--IQC for $\Del$ in \eqref{eq:sys},   $ \wh{M},  Q, Q_0,  R, R_0 \cge 0$ and symmetric $P$ 
	such that 
	\eqa{\label{eq:robust_ioss}}{
		& \chi_{k+1}^\top  P  \chi_{k+1} \leq \rho^{2}  \chi^\top_k P \chi_k +  \|w_k\|_{Q_0}^2 +  \|w_k-\ti{w}_k\|_Q^2  \\
		&  +\|\wh{x}_k-\ti{x}_k\|_{R_0}^2 +  \|\ti{z}_k \|^2_{\wh{M}}  + \|y_k -\ti{y}_k\|_R^2-z_{k}^\top (M+\wh{M}) z_{k} ,  \\
		& P \succ \diag(0,\rho^{-2}Z), 
	}
hold  	with \red{$\chi_k:=\col(x_k-\ti{x}_k, \psi_k-\ti{\psi}_k, x_k, \psi_k)$} for all   $k\in \N_0$,  every trajectory \red{$(\ti{x}_k,\ti{w}_k,\ti{d}_k, \kappa(\wh{x}_k), \ti{v}_k,\ti{y}_k, \ti{\psi}_k, \ti{z}_k)^\infty_{k=0} \in (\Z \times \R^{n_\psi} \times \R^{n_z})^\infty$} of the series connection of \eqref{eq:sys_1}--\eqref{eq:sys_3}  and the filter 
	\eqa{\label{eq:auxilary_filter}}{
		\ti{\psi}_{k+1} &= A_\Psi \ti{\psi}_k +B_\Psi \col(\ti{v}_k, \ti{d}_k),   \\
		\ti{z}_k &= C_\Psi \ti{\psi}_k +D_\Psi \col(\ti{v}_k, \ti{d}_k),
	}
	as well as  every  \red{$(x_k,w_k,d_k,\kappa(\wh{x}_k),v_k,y_k,\psi_k, z_k)^\infty_{k=0} \in (\Z \times \R^{n_\psi} \times \R^{n_z})^\infty$}  of the connection of \eqref{eq:sys}  and \eqref{eq:pointIQC_filter}.}

\Blu{The robust detectability, i.e. \eqref{eq:robust_ioss} together with \eqref{eq:pointIQC},    indicates that if the disturbance $w_k$,  the error of disturbances $w_k-\ti{w}_k$, \todo{the error $\wh{x}_k-\ti{x}_k$,}  the error of output measurements $y_k-\ti{y}_k$, and the error of filter outputs $z_k-\ti{z}_k$ approach  zero as $k\to \infty$, then   the uncertain system~\eqref{eq:sys} driven by  $u_k=\kappa(\todo{\wh{x}_k})$  is stabilized to origin and its state $x_k$ converges to the state $\ti{x}_k$ of the certain system  \eqref{eq:sys_1}--\eqref{eq:sys_3} driven by the same $u_k$.  
Indeed, plugging \eqref{eq:pointIQC} in \eqref{eq:robust_ioss} yields 
	\aln{
		& \chi_{k+1}^\top  \ov{P}  \chi_{k+1} \leq \rho^{2}  \chi^\top_k \ov{P} \chi_k +  \|w_k\|_{Q_0}^2 + \|\ti{z}_k \|^2_{\wh{M}}-\|z_{k}\|^2_{\wh{M}}  \\
		&     + \|w_k-\ti{w}_k\|_Q^2 + \|y_k -\ti{y}_k\|_R^2 + \todo{\|\wh{x}_k-\ti{x}_k\|^2_{R_0}}, 
	}     
	with $\ov{P}:=P-\diag(0,\rho^{-2}Z) \succ 0$.   
	  }
\Blu{\remark{In the  MHE literature, there are two classical notions of detectability:  $L$-step observability  \cite{kang2006moving, alessandri2008moving, michalska2002moving}  and   incremental input-output-to-state stability (i-IOSS)  \cite{sontag1997output, allan2021nonlinear}. All these notions consider  the incremental property of a specific dynamical  system.  When the explicit expression of the true system is assumed to be available and can be incorporated as a model into MHE, these notions are crucial for estimator design due to the link between incremental properties and estimation errors.   
In our problem setting, however,   only part of the explicit expression of the  system can be used to formulate  MHE  due to  the non-parametric uncertainty $\Delta$.  To analyze the stability of estimation errors, it is therefore essential to propose a notion of detectability  considering the discrepancy between the trajectory of  the uncertain true system~\eqref{eq:sys}  and that  of the model~\eqref{eq:sys_1}--\eqref{eq:sys_3} employed in the MHE formulation rather than the incremental property of the true system. As a result, the  state difference  $x_k-\ti{x}_k$ associated with the true system~\eqref{eq:sys} and the model~\eqref{eq:sys_1}--\eqref{eq:sys_3} is subject to $\Delta(g(x_k, w_k))$ rather than $\Delta(g(x_k, w_k))-\Delta(g(\ti{x}_k, \ti{w}_k))$, and hence is affected by the non-incremental  property of the system~\eqref{eq:sys}. 
Moreover, in the absence of $w_k$,  if $x_k$ approaches zero,  the uncertainty $\Delta$ will become increasingly negligible.  
These insights,   along with the intertwinement between the controlled system and MHE as indicated in Fig.~\ref{fig:closed_loop},  inspire us to incorporate the stability condition for the controlled uncertain system  into the notion of detectability, thereby justifying  considering both $x_k$ and $x_k-\ti{x}_k$ in \eqref{eq:robust_ioss}.
}}



\Blu{If  $\Delta$ satisfies point-wise $\rho$-IQC  incrementally, that is, 
\eql{\label{eq:incremenal_IQC}
e_{z,k}^\top M e_{z,k} -e_{\psi, k}^\top  Z e_{\psi, k}  +  \rho^{-2} e_{\psi, k+1}^\top  Z e_{\psi, k+1} \geq 0  
}
with $e_{\diamond,k}= \diamond_k -\ti{\diamond}_k$, $\diamond \in \{z, \psi\}$  for all $k \in \N_0$, where $z_k$ and $\ti{z}_k$ are the output of filter~$\Psi$ defined by the state-space realization $(A_\Psi, B_\Psi, C_\Psi, D_\Psi)$ with the input $\col(v_k, \Delta(v_k))$ and $\col(\ti{v}_k, \Delta(\ti{v}_k))$ respectively as well as  zero initial conditions. Then the proposed robust detectability in Definition~\ref{def:robust_detec} can be tailored to define a robust version of  i-IOSS for a class of  system. 
\definition{ [Robust i-IOSS] \label{def:robust_ioss_v2} 
	The system~\eqref{eq:sys}  is robustly i-IOSS if there exist $(M, Z, A_\Psi, B_\Psi, C_\Psi,D_\Psi, \rho)$  defining incremental point-wise $\rho$--IQC according to \eqref{eq:incremenal_IQC} for $\Del$,   $Q,   R \cge 0$ and symmetric $P$ such that 
	\eqa{\label{eq:robust_ioss_v2}}{
		& \bar{\chi}_{k+1}^\top  P  \bar{\chi}_{k+1} \leq \rho^{2}  \bar{\chi}^\top_k P \bar{\chi}_k +   \|w_k-\ti{w}_k\|_Q^2 + \|y_k -\ti{y}_k\|_R^2\\
        & -(z_{k}-\ti{z}_k)^\top M (z_{k}-\ti{z}_k),  \\
		& P \succ \diag(0,\rho^{-2}Z), 
	}
hold with $\bar{\chi}_k:=\col(x_k-\ti{x}_k, \psi_k-\ti{\psi}_k)$ for all   $k\in \N_0$,  all trajectories $(\ti{x}_k,\ti{w}_k,\ti{d}_k, u_k, \ti{v}_k,\ti{y}_k, \ti{\psi}_k, \ti{z}_k)^\infty_{k=0}$ and $(x_k,w_k, d_k, u_k, v_k,y_k, \psi_k, z_k)^\infty_{k=0}$  of the series connection of the system \eqref{eq:sys} and the filter $\Psi$ defined by $(A_\Psi, B_\Psi, C_\Psi, D_\Psi)$ with zero initial conditions.   
}
The above definition states that,  for any $\Delta$ satisfying incremental point-wise $\rho$-IQC,  the corresponding  system~\eqref{eq:sys} in series connection with the filter $\Psi$ fulfills
$$ \bar{\chi}_{k+1}^\top  \bar{P}  \bar{\chi}_{k+1} \leq \rho^{2}  \bar{\chi}^\top_k \bar{P} \bar{\chi}_k +   \|w_k-\ti{w}_k\|_Q^2 + \|y_k -\ti{y}_k\|_R^2  $$
 with some $\bar{P} \succ 0$, and hence is i-IOSS. 
This allows us to use the standard MHE design approach, e.g.\cite{Schiller2023lyapunov}, to construct a cost function  for a group  of systems, rather than  individually for each system,  provided that the mathematical expression of the system is precisely known. }  

\section{Verification of Detectability} \label{sec:verifiation}
\Blu{This section is dedicated to the numerical verification of the proposed detectability from Definition~\ref{def:robust_detec}. As a key tool for  the verification, we present the following lemma,  which modifies \cite[Lemma~7]{zemouche2013lmi} for a function $\Psi$ defined on its domain and codomain of different dimensions. } 
\Blu{
\lemma{ \label{lem:lipschitz_formulate}
The function $\Phi: \R^n \to \R^m$ is Lipschitz continuous on $\ov{\X}:=\X_1 \times \ldots \times \X_n \subeq \R^n$ with $\X_i \subeq \R$ and $i\in \I_{[1,n]}$, i.e., there exists $\gamma \geq 0$ such that 
\eql{ \label{eq:lem_lipschitz}
\|\Phi(x)-\Phi(y)\| \leq \gamma \|x-y\|, \ \forall x, y \in \ov{\X},  }
if and only if there exist functions $\phi_{ij}: \R^n \times \R^n \to \R$ 
and constants $\gamma_{ij, \min}$ and $\gamma_{ij, \max}$, so that $\forall x, y \in \ov{\X}$, 
\eql{
\Phi(x)-\Phi(y)=\sum^{m}_{i=1} \sum^{n}_{j=1} \phi_{ij}(x^y_{j-1}, x^y_{j}) e_{m}(i)e^\top_{n}(j)(x-y)
}
and 
\eql{\label{eq:lemma_bound}
\gamma_{ij,min} \leq \phi_{ij}(x^y_{j-1}, x^y_{j}) \leq \gamma_{ij,max},}
where  $x^y_j$ denotes $\col(\smat{I_j & 0}y, \smat{0 & I_{n-j}}x)$ and 
$e_m(i)$ denotes the $i$-th standard unit vector in $\R^m$. 
}
\proo{The main part of the proof parallels  \cite[Lemma~7]{zemouche2013lmi}. To show the necessity, we need to additionally ensure that $x^y_j \in \ov{\X}$ for any $j\in \I_{[1,n]}$ and any $x,y\in \ov{\X}$, which is always true for $\ov{\X}$ being a Cartesian  product of intervals. This permits invoking Lipschitz condition \eqref{eq:lem_lipschitz} to derive \eqref{eq:lemma_bound}.  }
}

If $(M,Z,A_\Psi, B_\Psi, C_\Psi, D_\Psi, \rho)$ defining  point-wise $\rho$--IQC is chosen in advance, then the verification of   robust detectability amounts to checking the condition \eqref{eq:robust_ioss}. To verify   \eqref{eq:robust_ioss},   let us first consider
\todo{\eqa{\label{eq: error_state}}{
	x_{k+1}-\ti{x}_{k+1} &= f(x_k,w_k, d_k, \kappa(\wh{x}_k)) -f(\ti{x}_k, \ti{w}_k,  \ti{d}_k,  \kappa(\wh{x}_k)), \\
	y_{k}-\ti{y}_k &= h(x_k, w_k, d_k, \kappa(\wh{x}_k))-h(\ti{x}_k, \ti{w}_k,  \ti{d}_k, \kappa(\wh{x}_k))\\
    v_k -\ti{v}_k &= g(x_k, w_k) - g(\ti{x}_k, \ti{w}_k). 
}} 
By noting that $f$, $h$, $g$ and $\kappa$ are  Lipschitz continuous and applying Lemma~\ref{lem:lipschitz_formulate}, we can   find  vectors $ \gamma_{1,min}, \gamma_{1,max}\in \R^{n_1}$, $\gamma_{2,min}, \gamma_{2,max}\in \R^{n_2}$, and $\gamma_{3,min},  \gamma_{3, max} \in \R^{n_3}$ with $n_1:=n(2n+n_w+p)$, $n_2:=m(2n+n_w+p)$ and \Blu{$n_3:=q(n+n_w)$},  matrix-valued  linear maps $A:   \R^{n_1} \to \R^{n \times n}$, $B_w:   \R^{n_1} \to \R^{n \times n_w}$,  $B_d:  \R^{n_1} \to \R^{n \times q}$, $C: \R^{n_2} \to \R^{m \times n}$, $D_w: \R^{n_2} \to \R^{m \times n_w}$,  $D_d: \R^{n_2} \to \R^{m \times q}$, \Blu{$C_v: \R^{n_3} \to \R^{q \times n}$,  and $E_{w}:\R^{n_3} \to \R^{q \times n_w}$}  such that \eqref{eq: error_state}  can be  rewritten as
\Blu{\eqa{ \label{eq:LPV_sys}}{
	e_{x,k+1}&=A(\Theta_1) e_{x,k} + B_{w}(\Theta_1)e_{w,k} + B_d(\Theta_1)e_{d,k},\\
	e_{y,k}&=C(\Theta_2) e_{x,k} + D_{w}(\Theta_2)e_{w,k} + D_d(\Theta_2)e_{d,k} \\
    e_{v,k} &=C_v(\Theta_3)e_{x,k} + E_{w}(\Theta_3)e_{w,k}
    }}
with the shorthand \red{$e_{\diamond,k}:= \diamond_k- \ti{\diamond}_k, \diamond \in \{x, w, d, y, v\}$},  some $\Theta_1 \in \mathbb{H}_1:= \{  \omega \in \R^{n_1}:  \gamma_{1,min} \leq  \omega \leq  \gamma_{1, max} \}$,  $\Theta_2 \in \mathbb{H}_2:= \{  \omega \in \R^{n_2}:  \gamma_{2,min} \leq  \omega \leq  \gamma_{2, max} \}$, \Blu{and $\Theta_3 \in \mathbb{H}_3:= \{  \omega \in \R^{n_3}:  \gamma_{3,min} \leq  \omega \leq  \gamma_{3, max}\}$}.
Recalling that $\Delta(0)=0$ and that the system~\eqref{eq:sys} controlled by $u_k=\kappa(x_k)$ satisfies  \eqref{eq:ISS_nominal}, we obtain
\Blu{$$f(0, 0, 0, \kappa(0))=0.$$} 
This enables us to reformulate \eqref{eq:sys_1} with \red{$u_k=\kappa(\wh{x}_k)$}  into
\eql{ \label{eq:LPV_sys_2}
	x_{k+1} =  A(\Theta_4) x_k + B_{w}(\Theta_4)w_k  + \todo{B_{u}}(\Theta_4)\todo{\wh{x}_k} + B_d(\Theta_4)d_k } 
 with some $\Theta_4 \in  \mathbb{H}_1$ and some linear matrix-valued maps $\todo{B_u}: \R^{n_1} \to \R^{n\times n}$, \red{similarly as} with \eqref{eq: error_state}. \Blu{By recalling that $g(0,0)=0$, we have 
 \eqa{}{
v_k &= C_v(\Theta_5) x_k + E_{w}(\Theta_5) w_k \\ 
\ti{v}_k & = C_v(\Theta_6) \ti{x}_k + E_{w}(\Theta_6) \ti{w}_k 
 }
with some $(\Theta_5, \Theta_6) \in \mathbb{H}_3 \times \mathbb{H}_3$.} 
We then define  \todo{$\nu_k:= \col(e_{w,k}, d_k, \ti{d}_k, w_k, \wh{x}_k )$,  $\zeta_k:=\col(e_{w,k}, w_k, e_{y,k}, \wh{x}_k-\ti{x}_k,  z_k, \ti{z}_k)$} to construct the \todo{extended} system
\eqa{\label{eq:sys_full}}{
	\chi_{k+1} &= \Ac(\Theta) \chi_k + \Bc(\Theta) \nu_k, \\
	\zeta_k &= \Cc(\Theta) \chi_k + \Dc(\Theta) \nu_k,
}
with  $ \bigl(\Ac(\Theta)$, $\Bc(\Theta)$, $\Cc(\Theta)$,  $\Dc(\Theta)\bigr)$ specified in \eqref{eq:sys_full_matrix} and  \Blu{$\Theta:=(\Theta_1, \Theta_2, \Theta_3, \Theta_4, \Theta_5, \Theta_6)$ in the (compact) box $\mathbb{H}:= \mathbb{H}_1 \times \mathbb{H}_2 \times \mathbb{H}_3 \times \mathbb{H}_1 \times \mathbb{H}_3 \times \mathbb{H}_3$}. 	
	
\prop{\label{prop:LMI_IQC}
	Given $(M, Z, A_\Psi, B_\Psi, C_\Psi,D_\Psi, \rho)$ defining $\rho$--IQC for $\Del$,  the condition~\eqref{eq:robust_ioss} holds if $\wh{M},  Q, Q_0,  R, \todo{R_0} \cge 0$,   $P \cg \diag(0, \rho^{-2}Z)$,  and  
	\eqa{\label{eq:robust_detec_LMI}}{
		(\bullet)^\top \mat{ccc}{ -\rho^{2}P &   & \\   &  P & \\ &  &  -P_p}\mat{cc}{I & 0\\ \Ac(\Theta) & \Bc( \Theta) \\ \Cc(\Theta) & \Dc(\Theta)  } \cle 0 
	}
	with  $P_p:=\diag(Q, Q_0, R, \todo{R_0},  -M-\wh{M}, \wh{M})$ for all $\Theta \in  \mathbb{H}$.  
}
\proo{By multiplying  \eqref{eq:robust_detec_LMI} from both sides by $\col(\chi_k, \nu_k)$ and its transpose and invoking  \eqref{eq:sys_full}, we get \eqref{eq:robust_ioss}. }
\Blu{\remark{Compared with other LMI-based methods \cite{arezki2023lmi, schiller2023moving} for the verification of i-IOSS-type detectability, our method is computationally less attractive, as the condition~\eqref{eq:robust_detec_LMI} needs to be validated over  a  higher dimensional  $\mathbb{H}$ in general. Moreover,   as we ignore the interdependence among elements in $\Theta$ by bounding them with boxes, the verifcation could be quite conservative  when the dimension $\mathbb{H}$ is very high,  However,  our method does not require  the system dynamics  to be differentiable. More importantly, as indicated by Lemma~\ref{lem:lipschitz_formulate},  the set $\mathbb{H}$ is subject to  the Lipschitz condition of system dynamics, the domain of system trajectory   does not therefore need to be bounded.  }}


To reduce conservatism  in the verification of  robust detectability, it is desirable not to fix all parameters of $\rho$--IQC but to treat at least some of them as free variables. In the sequel, we fix $(A_\Psi, B_\Psi, C_\Psi, D_\Psi)$ and $\rho\in(0,1)$, and then characterize families  of variables $M, Z$ for two  uncertainty classes, namely slope-restricted nonlinearities and  parametric uncertainties. This enables a joint verification of  $\rho$--IQC and  \eqref{eq:robust_ioss} via parameter dependent LMI  conditions.

\subsection{Slope restricted Nonlinearity }
Let $\Del(v_k)= \varphi(v_k)$, where $\varphi: \R^p \to \R^p$ satisfies
\eql{\label{eq:slope_restricted_0}
	\varphi(0)=0,
}
\eql{\label{eq:slope_restricted}
	\alpha \|x-y\|^2 \leq (\varphi(x)-\varphi(y))^\top (x-y) \leq \beta \|x-y\|^2 }
 for $x, y \in \R^p$ with the fixed constants  $\alpha,\beta \in \R$, $\alpha<\beta$.   The following lemma modifies \cite[Theorem~3]{morato2023stabilizing} for $\rho < 1 $ and the multi-variable  slope restricted condition~\eqref{eq:slope_restricted}. 
\lemma{ \label{lem:FIR-ZamesFalb}
	$\Del$ satisfies point-wise $\rho$--IQC w.r.t. $\rho\in (0,1)$,
	\aln{ &M=\smat{0 & W \otimes I_p  \\ W^\top \otimes I_p & 0}, \ Z=\smat{0 & Q \otimes I_p \\ Q^\top \otimes I_p & 0}\\
   	&A_\Psi= I_2 \otimes J_\nu \otimes I_p, \  B_\Psi = (I_2 \otimes \col(0_{ (\nu-1) \times 1},1) \otimes I_p)T\\
   	& C_\Psi= I_2 \otimes \col(I_\nu,0_{ 1 \times \nu }) \otimes I_p,  
   } 
   and $D_\Psi = (I_2 \otimes \col(0_{\nu \times 1}, 1) \otimes I_p)T$, where $J_\nu \in \R^{\nu \times \nu} $ is a Jordan block with eigenvalue $0$ and $T=\smat{\beta I_p & -I_p \\ -\alpha I_p  & I_p}$, if  
	$$
	\ov{W}:=W- \diag(Q,0) + \diag(0, \rho^{-2} Q) \in \S^{\nu+1} 
	$$
	is doubly hyperdominant, that is $\ov{W}_{ij} \leq 0$, for all $i\neq j $, $\sum^{\nu+1}_{j=1}\ov{W}_{ij}\geq 0$ for each $i$ and $\sum^{\nu+1}_{i=1}\ov{W}_{ij}\geq 0$  for each $j$. 
}
\proo{
Let us define $\varphi^\beta(v_k):= \beta v_k -\varphi(v_k)$, $\varphi^\alpha (v_k):= \varphi(v_k)-\alpha v_k$ and $\tilde{\varphi}^{\diamond}(v_k):=\col(\varphi^\diamond (v_{k-\nu}), \ldots, \varphi^\diamond (v_{k}) )$ for $\diamond \in \{\alpha, \beta\}$. Then  $z_k$ and $\psi_k$ generated by  $\Psi$ with$\psi_0=0$ and any input trajectory  $(\col( v_i, \varphi(v_i)))^{\infty}_{i=0}$ are 
\Blu{\aln{z_k &= \col(\tilde{\varphi}^\beta(v_k), \tilde{\varphi}^\alpha(v_k)), \\ 
		\psi_k&=(\varphi^\beta(v_{k-\nu}), \ldots, \varphi^\beta(v_{k-1}), \varphi^\alpha(v_{k-\nu}), \ldots, \varphi^\alpha(v_{k-1})).}}
Hence, the left side of inequality \eqref{eq:pointIQC} is reduced to 
\eql{ \label{eq:FIR_proof}
	(\tilde{\varphi}^\beta (v_k))^\top (\ov{W}\otimes I_p)\tilde{\varphi}^\alpha (v_k). }
The inequality \eqref{eq:slope_restricted} implies that the primitives of functions $\varphi^\alpha $ and $\varphi^\beta$ are  convex  by  \cite[Lemma~1]{zhou2018fenchel}. Henceforth, the primitives of $\tilde{\varphi}^\alpha $ and $\tilde{\varphi}^\beta$ are also convex. Moreover,    $\tilde{\varphi}^\alpha(0)=\tilde{\varphi}^\beta(0)=0$ due to \eqref{eq:slope_restricted_0}. We can hereby  apply \cite[Corollary~6]{scherer2023robust} to show that \eqref{eq:FIR_proof} is nonnegative if $\ov{W}$ is doubly hyperdominant, which finishes the proof.
}
\Blu{\remark{Since the parameter $\nu$ can be chosen freely, we can increase $\nu$ to increase the size of  variables $M$ and $Z$, thereby reducing conservatism in verifying detectability.}}   


Additionally, if $\varphi(x)=\col(\varphi_1(x_1), \ldots, \varphi_p(x_p))$ with $x=\col(x_1, \ldots, x_p)$ and $\varphi_i: \R \to \R$ satisfying \eqref{eq:slope_restricted} for each $i\in \I_{[1,p]}$, \Blu{then each $\varphi_i$ is also sector bounded by $[\alpha, \beta]$, i.e., 
$$(\varphi_i(x_i)-\alpha x_i)(\beta x -\varphi(x_i)) \geq 0, \  \forall x \in \R. $$
Hence, we can express $\varphi_i(x_i)=\delta_i x_i$ with some $\delta_i \in [\alpha, \beta]$.} This together with  
the arguments \Blu{regarding polytopic bounding} in \cite[Section~6.3.1]{fetzer2017classical} leads immediately to the following \red{result:} 
\lemma{\label{lem:polytope_iqc}$\Del$ satisfies point-wise $\rho$--IQC w.r.t. $\rho\in (0,1)$, 
	$M\in \S^{2p}$,  $Z \in \R$ and $A_\Psi=0$, $B_\Psi=0_{1 \times 2p}$, $C_\Psi=0_{2p \times 1}$ and $D_\Psi=I_{2p}$,
	if
	\eql{ \label{eq:poly_multiplier}
		(\bullet)^\top M \mat{c}{ I_p \\ \diag(\delta_1, \ldots, \delta_p) } \cge 0  }
		for all $\delta_i \in [\alpha, \beta]$ with $i=1,\ldots,p$.
}
\Blu{\proo{Multiplying \eqref{eq:poly_multiplier} by $\col(v_k)$ and its transpose implies 
$$(\bullet)^\top M \col(v_k, \varphi(v_k)) \geq 0, \forall k.$$
Since $z_k=\col(v_k, \varphi(v_k))$, we obtain \eqref{eq:pointIQC} for any $\rho$.}}

Since \eqref{eq:slope_restricted} holds trivially when $\varphi$ satisfies \eqref{eq:slope_restricted} component-wisely,  combining the results from Lemmas~\ref{lem:polytope_iqc} and \ref{lem:FIR-ZamesFalb} by adding the corresponding \eqref{eq:pointIQC} together may capture  the nature of uncertainties  with reduced conservatism. 

\subsection{ Parametric Uncertainty}
Let $\Del(v_k)= \delta I v_k \in \R^p$ with $\delta \in [a, b]$ and the fixed constants $a, b \in \R$, $a<b$. 
\lemma{\label{lem:fullblock_IQC} $\Del$ satisfies pointwise $\rho$--IQC w.r.t. $\rho \in (0,1)$, 
	
\aln{&M=\smat{M_1 & W_2 \\ W^\top_2 & M_3}, Z=\smat{Z_1 & Q_2 \\ Q^\top_2 & Z_3}, A_\Psi= I_2 \otimes A_\Phi, \\
	 &B_\Psi= (I_2 \otimes B_\Phi)T, 
	C_\Psi = (I_2 \otimes C_\Phi), D_\Psi = (I_2 \otimes D_\Phi)T}
     with $T=\smat{b I & -I \\ -a I & I}$, if 
	\eqa{\label{eq:multipler_para}}{
		(\bullet)^\top & \mat{ccc}{ -Z_i & & \\ & \rho^{-2}Z_i & \\ &  & M_i }\mat{cc}{ I & 0 \\ A_\Phi & B_\Phi \\ C_\Phi & D_\Phi} \cge 0 
	}
	for $i\in \{1,2,3\}$ with $M_2:= W_2+W^\top_2$ and $Z_2:=Q_2 + Q^\top_2$. 
}

Note that the above result  applies readily to the time-varying  $\delta_k \in [a, b]$ by fixing  \Blu{$(A_\Phi, B_\Phi,  D_\Phi)=(0,0,I)$   and following the similar reasoning as in the proof of Lemma~\ref{lem:fullblock_IQC}.}
For non-repeated time-varying uncertainties,  e.g., $\Delta(v_k)=\diag(\delta_{1,k}, \ldots, \delta_{p,k}) v_k$, $\delta_{i,k} \in [a, b]$, we can use  Lemma~\ref{lem:polytope_iqc} for this class of uncertainty by \Blu{following the same reasoning in the proof of Lemma~\ref{lem:polytope_iqc}.}     


\section{MHE-based Robust Stabilization}
\label{sec:MHE}

 Under the assumption that the system~\eqref{eq:sys} with the controller \eqref{eq:control_input} is robustly detectable according to Definition~\ref{def:robust_detec},  we propose a robust MHE scheme using the  past control inputs $u_i=\kappa(\wh{x}_i)$ and past  output measurements $y_i$ with $i \in \I_{[k-N_k, k-1]}$,  $N_k:= \min(k, N)$ and the estimation horizon  $N \in \N$ to estimate the  state $x_k$ at each time $k\in \N_0$.
To  account for $\Del$ in the MHE design,   we compute the estimate $\wh{\theta}_k:=\col(\wh{x}_k, \wh{\psi}_k)$  of the augmented state $\theta_k:= \col(x_k, \psi_k)\in \R^{n+n_{\psi}}$ associated with  the series connection of system~\eqref{eq:sys} and  the filter \eqref{eq:auxilary_filter}. Given the initial guess $\wh{x}_0$,  the estimate $\wh{\theta}_k$ is determined  by
\eqa{\label{eq:est_rule}}{
	\wh{\theta}_k & = \wh{\theta}^\star_{k|k}  :=\col(\wh{x}^\star_{k|k}, \wh{\psi}^\star_{k|k}),   \ k\in \N, \\
	\wh{\theta}_0 &= \col(\wh{x}_0, \wh{\psi}_0), \  \wh{\psi}_0=0,  }   
where \todo{$\wh{\theta}^\star_{k|k}$} is the minimizer to  the following optimization problem, 
\subeql{\label{eq:op_MHE}}{
	\min_{ \substack{\todo{\wh{\theta}_{\cdot|k}}, \wh{d}_{\cdot|k},   \wh{w}_{\cdot|k} } }   &  \ J( \todo{\wh{\theta}_{\cdot|k}},  \wh{w}_{\cdot|k},   \wh{y}_{\cdot|k}, \wh{z}_{\cdot|k}  )  \\
	\text{s.t.} 
	\  \wh{\theta}_{j+1|k} &= F(\wh{\theta}_{j|k},     \wh{w}_{j|k},  \wh{d}_{j|k}, \wh{x}_j), \label{eq:mhe_sys_1}  \\
	\mat{c}{  \wh{y}_{j|k} \\ \wh{z}_{j|k}} &= H(\wh{\theta}_{j|k},  \wh{w}_{j|k}, \wh{d}_{j|k}, \wh{x}_j),\label{eq:mhe_sys_2} \\
	\wh{w}_{j|k} \in  & \W,  \ \wh{y}_{j|k}\in \Y, \wh{x}_{j|k} \in \X,  \  j \in \I_{[k-N_k, k-1]} \\
	\wh{x}_{k|k} \in & \X,   \ \todo{\Lambda(\wh{\theta}_{\cdot|k}, \wh{w}_{\cdot|k}, \wh{y}_{\cdot|k} ) \leq 0},     
	 }
with functions $F$ and  $H$  defined by
\aln{
F(\wh{\theta}_{j|k}, \wh{w}_{j|k}, \wh{d}_{j|k}, \wh{x}_j)&:=\mat{c}{f(\wh{x}_{j|k}, \wh{w}_{j|k}, \wh{d}_{j|k}, \kappa(\wh{x}_j)) \\ A_\Psi \wh{\psi}_{j|k}+B_\Psi \smat{\Blu{g(\wh{x}_{j|k}, \wh{w}_{j|k})}\\ \wh{d}_{j|k} } },    \\
H(\wh{\theta}_{j|k}, \wh{w}_{j|k}, \wh{d}_{j|k},\wh{x}_j)&:=\mat{c}{h(\wh{x}_{j|k}, \wh{w}_{j|k}, \wh{d}_{j|k}, \kappa(\wh{x}_j)) \\ C_\Psi \wh{\psi}_{j|k}+D_\Psi \smat{\Blu{g(\wh{x}_{j|k}, \wh{w}_{j|k})}\\ \wh{d}_{j|k} } }.  
}
The cost function $J$ is given by 
\eqa{\label{eq:cost}}{
	 &J( \todo{\wh{\theta}_{\cdot|k}}, \wh{w}_{\cdot|k},  \wh{y}_{\cdot|k}, \wh{z}_{\cdot|k}  )  =\rho^{2N_k} (2+\todo{\eps}) \| \wh{\theta}_{k-N_k|k}- \wh{\theta}_{k-N_k}\|^2_{P_0}\\
	&+ \sum^{N_k}_{j=1} \rho^{2j-2} \left( (2 + \todo{\xi}) \|\wh{w}_{k-j|k}\|_Q^2 + \| \wh{z}_{k-j|k}\| ^2_{\wh{M}}   \right) \\
	& + \sum^{N_k}_{j=1} \rho^{2j-2} (\todo{1+\xi}) \|y_{k-j}-\wh{y}_{k-j|k} \|_R^2
	         }
\todo{and the constraint $\Lambda(\wh{\theta}_{\cdot|k},\wh{w}_{\cdot|k}, \wh{y}_{\cdot|k}) \leq 0$ is described by}
\eqa{\label{eq:constraint}}{
&\sum_{j=1}^{N_k} \rho^{2j-2}  \|\wh{x}_{k-j}-\wh{x}_{k-j|k}\|^2_{R_0} 
\leq \eps \rho^{2N_k}   \|\wh{\theta}_{k-N_k}-\wh{\theta}_{k-N_k|k}\|^2_{P_0}\\
& + \sum^{N_k}_{j=1} \xi \rho^{2j-2}   (\|\wh{w}_{k-j|k}\|_Q^2 + \|y_{k-j}-\wh{y}_{k-j|k} \|_R^2 ) , }
\todo{with some $\eps >0$, $\xi\geq 0$ and}   $P_0 \cge (\bullet)^\top P \col(I_{n+n_{\psi}}, 0)$, where  $Q$, $\wh{M}$, $R$, \todo{$R_0$}, $P$,  and $\rho$ are   specified in Definition~\ref{def:robust_detec}. 
\todo{Since $\wh{\psi}_{k-N_k|k}$ in $\wh{\theta}_{k-N_k|k}$ can be  chosen freely regardless of the constraints on $\wh{w}_{\cdot|k}$,  $\wh{y}_{\cdot|k}$ and $\wh{x}_{\cdot|k}$,   \eqref{eq:constraint} is always feasible. Further,  any trajectory of the true system~\eqref{eq:sys} is also a solution of \eqref{eq:sys_1}--\eqref{eq:sys_3}. Hence,  the problem~\eqref{eq:op_MHE} is always feasible.}
\Blu{\remark{  In contrast to the standard MHE formulation in \cite{schiller2023moving}, the above cost function~\eqref{eq:cost} contains the additional penalization term  $\|\wh{z}_{\cdot|k} \|^2_{\wh{M}}$.   This is attributed  to  the proposed notion of robust detectability  in Definition~\ref{def:robust_detec}, which involves  $\|\ti{z}_k\|^2_{\wh{M}}$.  
   The penalization term  $\|\wh{z}_{\cdot|k} \|^2_{\wh{M}}$ in  \eqref{eq:cost}  can be eliminated if we restrict $\wh{M}$ to be zero matrix in \eqref{eq:robust_ioss}. However, this may result in infeasibility in the  verification of  detectability. Actually, the weight $\wh{M}$ in Definition~\ref{def:robust_detec} can be relaxed to an indefinite matrix.  This will yet lead to nonconvex cost functions,  rendering it difficult to solve  the  problem~\eqref{eq:op_MHE}.      
}
}

\theorem{ \label{theo:closed_loop_stability} Assume that the system~\eqref{eq:sys} with the controller \eqref{eq:control_input} is robustly detectable according to Definition~\ref{def:robust_detec}. Let  the estimation horizon $N \in \N$ in \eqref{eq:op_MHE} be chosen such that 
	\eql{\label{eq:choice_N}
		N>-\log_{\rho^2}(\ov{\lambda}(P_2, P_1)),   
	}
	with $P_1:=P-\diag(0,\rho^{-2}Z)$ and 
    $$P_2:=P_1+ (\bullet)^\top P^{-1}_{11} \smat{ P_{11} & P_{12}} +\diag( (2+\todo{\eps})P_0, 0),$$
    where $P_{11}, P_{12} \in \R^{(n+n_{\psi})\times (n+n_{\psi})}$ are blocks of $P=\smat{P_{11} & P_{12} \\ P^\top_{12} &  P_{22}}$ chosen according to   Definition~\ref{def:robust_detec}.  	
	Then the closed-loop system formed of the system~\eqref{eq:sys} with the controller~\eqref{eq:control_input} and the MHE described by \eqref{eq:est_rule}  and  \eqref{eq:op_MHE} is ISS, that is, there exist $\wh{\beta} \in \Kc \Lc$ and $\wh{\alpha} \in \Kc$ such that \eqref{eq:RGES} holds for all $k\in \N_0$.  
}
\proo{ The core of  the proof is the construction of the so-called $M$-step Lyapunov function from \cite{Schiller2023lyapunov} \Blu{despite of the uncertainty $\Delta$. This is enabled by leveraging the proposed IQC-based robust detectability in Def.~\ref{def:robust_detec}. }  

The second inequality in \eqref{eq:robust_ioss} implies $P_{11}\succ 0$.  By invoking Schur complement and noting that  $P_{11}\succ 0$, we have
$$
\mat{ccc}{P_{11} & P_{11}  & P_{12}\\ P_{11} & P_{11} & P_{12} \\ P^\top_{12} & P^{\top}_{12} & P_{22}} - \diag(2P_{11}, \wh{P}) \cle 0 
$$
with $\wh{P}:= P+ (\bullet)^\top P^{-1}_{11} \smat{ P_{11} & P_{12}}$.
Multiplying this inequality by $\col(z,x,y)$ and its transpose from both sides yields    
\eql{\label{eq:proof_matix_ineq} 
	(\bullet)^\top \mat{cc}{P_{11} & P_{12} \\ P^\top_{12} & P_{22}} \mat{c}{x+z \\ y} \leq (\bullet)^\top \wh{P} \mat{c}{x \\ y} + 2 \|z\|^2_{P_{11}}  
}
for all $x, y, z$.  

Let us define  $\chi_{k|j}:=\col(\theta_k-\wh{\theta}^\star_{k|j}, \theta_k)$ and apply the first inequality in  \eqref{eq:robust_ioss} successively  to get 
\eqa{\label{eq:proof_bound_1}}{
	& \chi^\top_{k|k} P \chi_{k|k} \leq \sum^{N_k}_{j=1} \rho^{2j-2} \left( 2\|\wh{w}^\star_{k-j|k}\|^2_{Q} +   \|w_{k-j} \|^2_{2Q+Q_0} \right) \\
	& + \sum^{N_k}_{j=1} \rho^{2j-2} \left(   \|y_{k-j}-\wh{y}^\star_{k-j|k} \|^2_R + \todo{\|\wh{x}_{k-j}-\wh{x}^\star_{k-j|k} \|^2_{R_0}}    \right) \\
	& - \sum^{N_k}_{j=1}  \rho^{2j-2} \left( z^\top_{k-j} (M+\wh{M}) z_{k-j}-\|\wh{z}^{\star}_{k-j|k}\|^2_{\wh{M}} \right) \\
	& + \rho^{2N_k}   \chi^\top_{k-N_k|k} P \chi_{k-N_k|k}. 
}
By leveraging  \eqref{eq:proof_matix_ineq} and  \eqref{eq:est_rule}, we obtain 
\eqa{\label{eq:proof_bound_2}}{  \chi^\top_{k-N_k|k} P \chi_{k-N_k|k} & \leq \chi^\top_{k-N_k|k-N_k} \wh{P} \chi_{k-N_k|k-N_k}\\
	&  + 2 \|\wh{\theta}_{k-N_k}-\wh{\theta}^{\star}_{k-N_k|k} \|^{\todo{2}}_{P_{11}}.}
Let $\chi_k:=\chi_{k|k}$.  By inserting \eqref{eq:proof_bound_2}, \todo{\eqref{eq:constraint}} and \eqref{eq:cost} into \eqref{eq:proof_bound_1} as well as noting that $P_0 \cge P_{11}$,  we get 
\eqa{\label{eq:proof_bound_3}}{
	& \chi^\top_{k} P \chi_k \leq  \rho^{2N_k}   \chi^\top_{k-N_k} \wh{P} \chi_{k-N_k}  + J(\wh{\theta}^\star_{\cdot|k}, \wh{w}^\star_{\cdot|k},   \wh{y}^\star_{\cdot|k}, \wh{z}^\star_{\cdot|k}) \\
	&+\sum^{N_k}_{j=1} \rho^{2j-2}  (\|w_{k-j}\|^2_{\todo{2Q+Q_0}}  -z^\top_{k-j}(M+\wh{M}) z_{k-j} ).  
}
By optimality, i.e., 
\mun{
	J(\wh{\theta}^\star_{\cdot|k}, \wh{w}^\star_{\cdot|k},  \wh{y}^\star_{\cdot|k}, \wh{z}^\star_{\cdot|k}) \leq \rho^{2N_k}  \| \theta_{k-N_k}-\wh{\theta}_{k-N_k}\|^2_{\todo{(2+\eps)}P_0} \\
	+\sum^{N_k}_{j=1}\rho^{2j-2}((2+\xi)\|w_{k-j}\|^2_Q+\|z_{k-j}\|^2_{\wh{M}}), 
}
and \eqref{eq:est_rule}, we derive from \eqref{eq:proof_bound_3} that
\eqa{\label{eq:proof_M-step}}{
	& \chi^\top_k P \chi_k   \leq \sum^{N_k}_{j=1} \rho^{2j-2} ( \|w_{k-j}\|^2_{\todo{(4+\xi)Q+Q_0}}  -z^\top_{k-j}Mz_{k-j} )   \\
	&   + \rho^{2N_k}   \chi^\top_{k-N_k} \bigl(\wh{P}+\diag((\todo{2+\eps})P_0, 0) \bigr) \chi_{k-N_k}.      
}
By  multiplying  \eqref{eq:pointIQC} by $\rho^{-2k-2}>0$ and then summing it from $k = \bar{k}-N_k$ to $ \bar{k}-1$ with  $\bar{k}>N_k$,  we get 
\mun{
	\rho^{-2\bar{k}-2} \psi^\top_{\bar{k}} Z \psi_{\bar{k}} - \rho^{-2(\bar{k}-N_k+1)} \psi^{\top}_{\bar{k}-N_k} Z \psi_{\bar{k}-N_k}
	\\
	\sum^{N_k}_{j=1} \rho^{-2(\bar{k}-j+1)}z^\top_{\bar{k}-j}Mz_{\bar{k}-j}\geq 0.
}
As a result, we have 
\equ{
	\sum^{N_k}_{j=1} \rho^{2j-2}z^\top_{k-j}Mz_{k-j} \geq \rho^{2N_k-2} \psi^{\top}_{\bar{k}-N_k} Z \psi_{\bar{k}-N_k}- \rho^{-2} \psi^\top_{k} Z \psi_{k}.
}
Combining this with \eqref{eq:proof_M-step} leads to
\eqa{\label{eq:proof_M-step_3}}{
	& \chi^\top_k \underbrace{(P-\diag(0,\rho^{-2}Z))}_{=P_1}  \chi_k \leq \sum^{N_k}_{j=1} \rho^{2j-2} \|w_{k-j}\|^2_{(4+\xi)Q+Q_0} \\
	& + \rho^{2N_k} \chi^\top_{k-N_k} \underbrace{ \bigl(\wh{P}+\diag((2\todo{+\eps})P_0,0, -\rho^{-2}Z) \bigr) }_{=P_2}  \chi_{k-N_k} .    
}
From the second inequality in \eqref{eq:robust_ioss} and by noting that $P_0 \cge P_{11} \cg 0$,  we have $P_1 , P_2 \succ 0$.  Applying $P_2 \leq \ov{\lambda}(P_2, P_1) P_1$ to \eqref{eq:proof_M-step_3} yields 
\eqa{\label{eq:proof_M-step_3_2}}{
	\| \chi_k\|^2_{P_1}  \leq \rho^{2N_k} \ov{\lambda}(P_2, P_1) \|\chi_{k-N_k}\|^2_{  P_1} 
	+ \sum^{N_k}_{j=1} \rho^{2j-2} \|w_{k-j}\|^2_{\wh{Q}}     
}
with $\wh{Q}:= (4+\xi)Q+Q_0$. 
Let us  define $\lambda_m:=\ov{\lambda}(P_2, P_1)$,  $\mu:=\rho \lambda_m^{1/(2N)}$, $\tau:= k- \lfloor k/N \rfloor$ and $ \lambda:= \max(\mu,\rho)$, where  $\lambda \in (0, 1)$ in view of \eqref{eq:choice_N}.  Then,  the inequality \eqref{eq:proof_M-step_3_2} implies 
\eqa{\label{eq:proof_M-step_4}}{
	& \|\chi_k\|^2_{P_1} \leq 
	\sum^{\lfloor k/N \rfloor -1}_{i=0} \mu^{2N} \sum^N_{j=1} \rho^{2j-2}\|w_{k-iN-j}\|^2_{\wh{Q}} \\
	& + \mu^{2\lfloor k/N \rfloor N} \Bigl( \rho^{2\tau}\lambda_m  \| \chi_{0}  \|^2_{P_1} + \sum^{\tau}_{j=1} \rho^{2j-2}\|w_{\tau-j}\|^2_{\wh{Q}}   \Bigr) \\
	& \leq \lambda^{2k} \lambda_m \|\chi_0\|^2_{P_1} + \sum^{k}_{i=1}\lambda^{2i-2} \|w_{k-i}\|^2_{\wh{Q}}.  }
We consider a permutation matrix 
$$T=\diag\Bigl(I_{n}, \smat{ 0 & I_{n} \\ I_{n_\psi} & 0 }, I_{n_\psi}\Bigr),  $$
and define $\wh{P}_1:=T P_1 T^\top$, where  $\wh{P}_1 \succ 0$ by $P_1 \succ 0$, to rewrite  \eqref{eq:proof_M-step_4}  into
\eqa{\label{eq:proof_M-step_6}}{
	\|T\chi_k\|^2_{\wh{P}_1} \leq \lambda^{2k} \lambda_m \|T \chi_0\|^2_{\wh{P}_1} + \sum^{k}_{i=1}\lambda^{2i-2} \|w_{k-i}\|^2_{\wh{Q}}   
}
with $T\chi_k = \col(x_k-\wh{x}_k, x_k, \psi_k -\wh{\psi}_k, \psi_k)$.  
From \eqref{eq:proof_M-step_6} and by noting that $\psi_0=\wh{\psi}_0=0$ given in \eqref{eq:pointIQC_filter} and \eqref{eq:est_rule}, we have 
\eqa{\label{eq:proof_M-step_7}}{
	\|T\chi_k\|^2_{\wh{P}_1} \leq \lambda^{2k} \lambda_m  \mam{c}{x_0-\wh{x}_0\\ x_0 }^2_{\wh{P}_{11}} + \sum^{k}_{i=1}\lambda^{2i-2} \|w_{k-i}\|^2_{\wh{Q}},   
}
where $\wh{P}_{11}$ is one block in $\wh{P}_1 = \smat{\wh{P}_{11} & \wh{P}_{12} \\ \wh{P}^\top_{12} & \wh{P}_{22}} \in \S^{2n+2n_{\psi}}$.  
Let us choose a matrix $X \cg 0$ such that 
$$X \cle \wh{P}_{11} -\wh{P}_{12}\wh{P}^{-1}_{22} \wh{P}^\top_{12},$$
 which is always possible due to $\wh{P}_1 \cg 0$.  Then $\diag(X, 0) \cle \wh{P}_1 $ by Schur complement.  Let us define  $c_x:= \ov{\lambda}(I_{2n}, X)$, $c_0:= \ov{\lambda}(\wh{P}_{11}, I_{2n})$ and $c_w:= \ov{\lambda}(\wh{Q},I_{n_w})$. From \eqref{eq:proof_M-step_7} and by applying the relation $\|a\|^2+\|b\|^2\leq (\|a\|+\|b\|)^2$,   we arrive at 
\eqa{\label{eq:proof_theo_final}}{
	\|x_k\| &\leq \lambda^{k} C_x \mam{c}{x_0-\wh{x}_0 \\ x_0} + \sum^{k}_{i=1}\lambda^{i-1} C_w \|w_{k-i}\|\\
	&\leq \lambda^{k}C_x(\|x_0-\wh{x}_0 \| + \|x_0\|) + \frac{C_w}{1-\lambda} \max_{i\in \I_{[0, k-1]}} \|w_i\|,  }
with $C_x:= \sqrt{\lambda_m c_x c_0}$ and $C_w:=\sqrt{c_x c_w}$.
This leads to \eqref{eq:RGES} by choosing $\wh{\beta}(r,k)=C_x\lambda^k r$ and $\wh{\alpha}(r)=(C_w r) /(1-\lambda)$, and hence completes the proof.  }

Following the same line of reasoning as in Theorem~\ref{theo:closed_loop_stability} and using the relation 
\cite[(25)]{knufer2020time},  we can show the boundedness (convergence) of estimation errors under bounded (convergent) disturbances, as stated in the sequel.  
\corollary{\label{coro:MHE_stability} Assume that the system~\eqref{eq:sys} with \eqref{eq:control_input} is robustly detectable. If the estimation horizon $N$ is chosen such that  \eqref{eq:choice_N} holds, then there exist $c_x, c_w > 0$  such that the estimate $\wh{x}_k$ determined by \eqref{eq:op_MHE} and \eqref{eq:est_rule} satisfies
	$$ 
	\|x_k-\wh{x}_k\|\leq \lambda^k c_x  \mam{c}{x_0\\x_0-\wh{x}_0} + c_w \todo{\max_{i\in \I_{[1, k]}} (\sqrt{\lambda}^{i-1} \|w_{k-i}\|)} 
	$$
	with  $\lambda=\rho \max( (\ov{\lambda}(P_2, P_1))^{\frac{1}{2N}}, 1)$  for all $k \in \N_0$, all initial conditions $ \wh{x}_0 \in \X$, and any trajectory $(x_i,w_i,d_i,\kappa(\wh{x}_i), v_i,y_i,)^\infty_{i=0} \in \Z^\infty$ of \eqref{eq:sys}. 
}
\remark{\Blu{Due to $P_{11} \succ 0$ and $P_0 \succ 0$, $P_2-P_1$ from Theorem~\ref{theo:closed_loop_stability} satisfies $P_2-P_1 \todo{\cge} 0$ and $P_2 \neq P_1$. Hence, we have   $\lambda=\rho (\ov{\lambda}(P_2, P_1))^{\frac{1}{2N}}$ with $\ov{\lambda}(P_2, P_1) > 1$. Therefore, the decay rate $\lambda \in (0,1)$ improves as $N$ increases, resulting in an enhanced estimation performance. Further, the improved decay rate together with \eqref{eq:proof_theo_final}   implies a faster stabilization of the closed-loop system.}  }

\remark{ Noting that $\ov{\lambda}(P_2, P_1)$ increase as $\eps$ grows, one should therefore choose small $\eps$  for small decay rate $\lambda$. If $\xi$ is also chosen to be small, then  $\wh{\psi}_{k-N_k|k}$ could be driven to a very large value due to constraint~\eqref{eq:constraint}, yielding possibly large $\wh{z}_{\cdot|k}$. Consequently, the cost function $J$ could be  dominated by the penalization term $\|\wh{z}_{\cdot|k}\|^2_{\wh{M}}$.  MHE then tends to ignore the output measurements and prior state estimate, resulting to degraded estimation performances. Therefore,  $\xi$ should be chosen to be large.      }



\section{Numerical Example}
\label{sec:example}

We illustrate the theoretical results by considering the following uncertain nonlinear system 
\aln{
	x_{1,k+1}&=1.3 x_{1,k} -0.4 x_{2,k} -d_k- 0.1 \sin(0.5 x_{1,k})+u_{1,k},  \\
	x_{2,k+1}&=0.6 x_{1,k} +0.75 x_{2,k}+u_{2,k}, \\
	y_k &= x_{2,k}+w_k, \ v_k= x_{1,k}, \ d_k = \Del(v_k),  
}
with the uncertainty $\Del(v)= 0.125 (| v+2 |-|v-2|)$ and the control inputs $(u_{1,k}, u_{2,k}) = (0.5 \wh{x}_{1,k} -0.41 \wh{x}_{2,k}, 0.4\wh{x}_{1,k} -0.75\wh{x}_{2,k})$. 
The disturbance $w_k$ is a uniformly distributed random variable satisfying $w_k \in \W = [-0.1, 0.1]$. The state $x_k$ and output $y_k$ are evolved in $\X=\R^2$ and $\Y=\R$ respectively for all $k\in \N_0$.   

Robust detectability in Definition~\ref{def:robust_detec}  is verified by \eqref{eq:robust_detec_LMI} together with $\rho^2=0.86$ and the set of   $(M,Z)$  from  Lemma~\ref{lem:polytope_iqc} combined with that from    Lemma~\ref{lem:FIR-ZamesFalb}  with $\nu=2$, $\beta=0.25$ and $\alpha=0$.  Note that \eqref{eq:robust_detec_LMI} only with  $(M,Z)$ from   Lemma~\ref{lem:polytope_iqc} is infeasible.   
We choose $P_0=(\bullet)^\top P \col(I_{6}, 0)$, $\eps=0.1$ and $\xi=500$. \Blu{The theoretical minimum estimation horizon computed from \eqref{eq:choice_N} is $N_{min}=12$.  As a benchmark for  the comparison, we implement the standard  MHE from \cite{Schiller2023lyapunov}  by ignoring the uncertainty $\Delta$, i.e., $d_k=0$ for all $k \in \N_0$. The minimum estimation horizon for standard MHE is $N_{min}=10$.     We choose  the estimation horizon $N = 15$ for the proposed and the standard MHE schemes.
}    
\begin{figure}[h]
	\centering
	\begin{subfigure}{0.21\textwidth}
		\vspace{5pt}
		\hspace{6pt}
		\includegraphics[width=0.96\textwidth, clip=false, trim = 20mm 6mm 15mm 20mm]{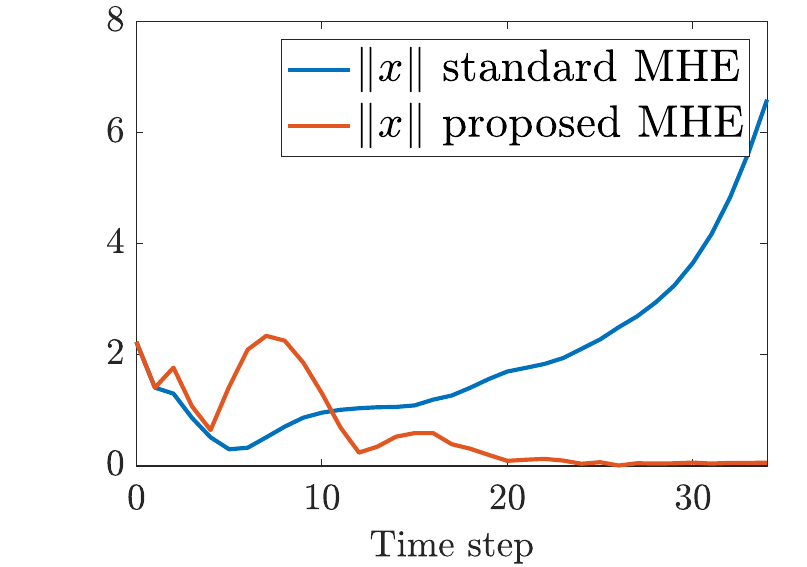}
		\caption{}
	\end{subfigure}
	\hspace{13pt}
	\begin{subfigure}{0.21\textwidth}
		\vspace{13pt}
		\hspace{-1pt}
		\includegraphics[width=0.96\textwidth, clip=false, trim = 20mm 6mm 15mm 20mm]{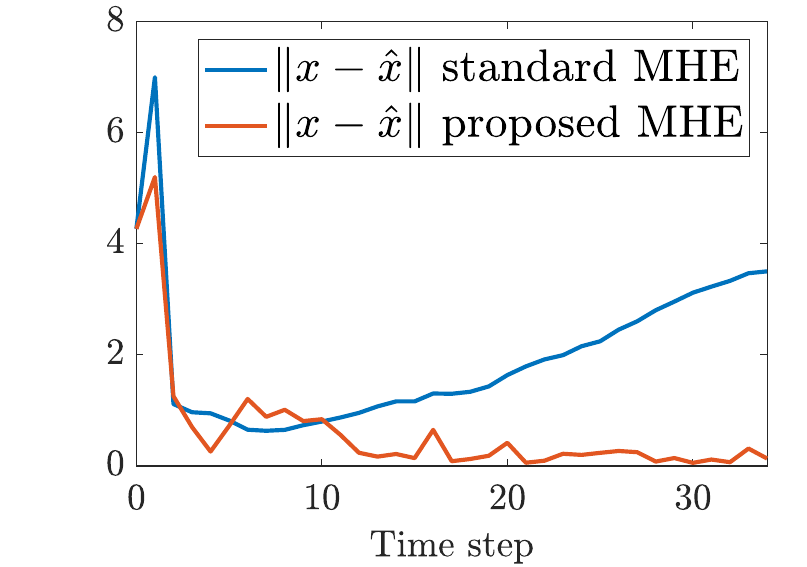}
		\caption{}
	     \end{subfigure}
	\vspace{-1pt}
	\caption{ Closed-loop trajectories and estimation errors}
	\label{fig:sim_plot}
\end{figure}

\Blu{As shown in Fig.~\ref{fig:sim_plot}, the controlled system using the proposed MHE  is effectively stabilized close to the origin, exhibiting negligible   estimation errors in the end,  when compared to the  standard MHE.  This clearly showcases the merit of the proposed MHE. }

\section{Conclusion}
\label{sec:conclusion}
A \red{practical} robust MHE scheme is proposed for the feedback control of \red{general} nonlinear constrained systems with possibly  nonlinear uncertainties. We introduced the concept of robust detectability by employing IQCs to  design MHE such that the uncertain closed-loop system remains ISS w.r.t. exogenous disturbances. As a possible extension, we will consider dynamical  uncertainties by employing  more general  IQCs, e.g., finite-horizon IQCs with a terminal cost from \cite{scherer2022dissipativity}.
\Blu{Moreover, future research will explore the integration of the proposed MHE scheme with more classes of controllers, e.g., dynamic state feedback  controller.}

\section*{Appendix}
The system matrices in the system~\eqref{eq:sys_full} are given by
\todo{\eqa{\label{eq:sys_full_matrix}}{
	& \Ac(\Theta) = \mats{cccc}{A(\Theta_1) & 0 & 0 & 0\\
		B_\Psi\smat{C_v(\Theta_3) \\ 0} & A_\Psi & 0 & 0\\
     	0  & 0 & A(\Theta_4) & 0\\
		0 & 0 & B_\Psi\smat{C_v(\Theta_5) \\ 0} & A_\Psi} \\
	& \Bc (\Theta)  = \mats{ccccc}{B_w(\Theta_1) & B_d(\Theta_1) &  -B_d(\Theta_1) & 0  & 0\\
	    B_\Psi \smat{E_{w}(\Theta_3) \\ 0}  & B_\Psi\smat{0 \\I}   & -B_\Psi\smat{0 \\I } & 0 & 0 \\
		0 & B_d(\Theta_4) & 0 & B_w(\Theta_4) & B_u(\Theta_4) \\
		0  & B_\Psi\smat{0 \\ I} & 0 &  B_{\Psi}\smat{E_{w}(\Theta_5) \\ 0 } & 0  }\\
	& \Cc(\Theta) = \mats{cccc}{0 & 0 & 0 & 0\\ 0 & 0 & 0 & 0 \\ C(\Theta_2) & 0 & 0 & 0\\
		I & 0 & -I & 0 \\	
		0 & 0 &  D_\Psi \smat{C_v(\Theta_5) \\ 0}  & C_\Psi  \\
		-D_\Psi \smat{C_v(\Theta_6) \\ 0} & -C_\Psi &  D_\Psi \smat{C_v(\Theta_6) \\ 0} & C_\Psi 
	}\\
	& \Dc(\Theta) = \mats{ccccc}{ I & 0 & 0 & 0 & 0 \\
		0 & 0 & 0 & I  & 0\\
		D_{w}(\Theta_2) & D_{d}(\Theta_2) & -D_{d}(\Theta_2) & 0 & 0 \\
			0 & 0 & 0 & 0 & I \\
		0  &  D_\Psi\smat{0 \\I} & 0  & D_\psi \smat{E_{w}(\Theta_5) \\0}  & 0\\
		-D_{\psi}\smat{E_{w}(\Theta_5)\\0} & 0 & D_\Psi \smat{0 \\ I} & D_{\psi}\smat{E_{w}(\Theta_5)\\0} & 0 
	  }. }}

\begin{proofnew}[Proof of Lemma~\ref{lem:fullblock_IQC}]
 Let us define $\col(\tilde{v}_k, \tilde{d}_k) :=T\col(v_k,d_k)=\col( (b-\delta)v_k, (\delta-a)v_k)$. Let $\phi_k$ and $\tilde{z}_k$ be the state and output of the filter $\Phi$ with  the  state-space realization $(A_\Phi, B_\Phi, C_\Phi, D_\Phi)$, the initial condition  $\phi_0=0$ and  the input $\tilde{v}_k$.  Multiplying the left side of \eqref{eq:multipler_para} by $\col(\phi_k, \tilde{v}_k)$ and its transpose leads to 
	\eql{ \label{eq:IQC_para_proof}
		\tilde{z}_k^\top M_i \tilde{z}_k -\phi_k^\top Z_i \phi_k +\rho^{-2}\phi^\top_{k+1} Z_i \phi_{k+1} \geq 0  } 
	for all $ \ i\in \{1,2,3\}$.   For $\delta \neq b$, we have $\tilde{d}_k= \tilde{\delta} \tilde{v}_k$ with $\tilde{\delta}=(\delta-a)/(b-\delta)>0$. By linearity of $\Phi$,  the left side of \eqref{eq:pointIQC} reads  
	\mun{\tilde{z}_k^\top (M_1+ \tilde{\delta} M_2+ \tilde{\delta}^2 M_3) \tilde{z}_k- \phi_k^\top (Z_1+ \tilde{\delta} Z_2+ \tilde{\delta}^2 Z_3) \phi_k \\
	 + \rho^{-2}\phi_{k+1}^\top (Z_1+ \tilde{\delta} Z_2+ \tilde{\delta}^2 Z_3) \phi_{k+1},}
	 which  is nonnegative due to \eqref{eq:IQC_para_proof}, and thereby shows that  \eqref{eq:pointIQC} is valid for $\delta \in [a,b)$. For $\delta=b$, we can consider $\phi_k$ and $\tilde{z}_k$ as the state and output of $\Phi$ driven by $\tilde{d}_k$ with $\phi_0=0$. Following the same reasoning as $\delta \neq b$, we can conclude that  \eqref{eq:pointIQC} is valid for $\delta \in [a, b]$.    
\end{proofnew}

\bibliographystyle{IEEEtran}
\bibliography{Ref}

\end{document}